\documentclass[aps,prb,twocolumn,showpacs]{revtex4}
\usepackage{epsfig}
\usepackage{times}
\usepackage{amsmath}
\begin{document}

\title{Stochastic resonance in soft matter systems: combined effects of static and dynamic disorder}

\author{Matja{\v z} Perc}
\email{matjaz.perc@uni-mb.si}
\author{Marko Gosak}
\email{marko.gosak@uni-mb.si}
\author{Samo Kralj}
\email{samo.kralj@uni-mb.si}
\affiliation{Department of Physics, Faculty of Natural Sciences and Mathematics, University of \\ Maribor, Koro{\v s}ka cesta 160, SI-2000 Maribor, Slovenia\\
$^\ddagger$Jo{\v z}ef Stefan Institute, P.O. Box 3000, SI-1000 Ljubljana, Slovenia}

\begin{abstract}
We study the impact of static and dynamic disorder on the phenomenon of stochastic resonance (SR) in a representative soft matter system. Due to their extreme susceptibility to weak perturbations soft matter systems appear to be excellent candidates for the observation of SR. Indeed, we derive generic SR equations from a polymer stabilized ferroelectric liquid crystal (LC) cell, which is a typical soft matter representative constituting one of the basic components in several electro-optic applications. We generalize these equations further in order to study an even broader class of qualitatively different systems, especially disclosing the influence of different types of static disorder and interaction ranges amongst LC molecules on the SR response. We determine the required conditions for the observation of SR in the examined system, and moreover, reveal that a random field type static disorder yields qualitatively different responses with respect to random dilution, random bond and spin glass universality classes. In particular, while the latter three decrease the level of dynamic disorder (Gaussian noise) warranting the optimal response, the former evokes exactly the opposite effect, hence increasing the optimal noise level that is needed to resonantly fine-tune the system's response in accordance with the weak deterministic electric field. These observations are shown to be independent of the system size and range of interactions, thus implying their general validity and potentially wide applicability also within other similar settings. We argue that soft matter systems might be particularly adequate as a base for different SR-based sensitive detectors and thus potent candidates for additional theoretical as well as experimental research in the presently outlined direction.
\end{abstract}

\maketitle

\section{Introduction}

When noise is introduced to nonlinear systems one can observe a variety of interesting and counterintuitive phenomena.\cite{Hor84, Han99, SA07} Perhaps the most prominent is the phenomenon of SR where an appropriate intensity of noise evokes the best correlation between a weak deterministic stimulus and the response of a nonlinear system.\cite{Sr1, Sr2, SrX, Sr3, Sr4, Sr5, Sr6, Sr7, Sr8, Sr9} This contradicts intuitive reasoning that suggests noise can only act destructive on system performance. To observe the phenomenon of SR a nonlinear systems must contain the following three basic ingredients: i) an energetic activation barrier (\textit{i.e.} a threshold), ii) a weak coherent input (\textit{i.e.} periodic external forcing), and iii) a source of noise. These conditions are very general indeed, enabling the observation of SR in the most diverse circumstances ranging from neuronal and brain functioning \cite{Br1, Br2, Br3, Br4, Brx, Br5, Br6} to the recurring occurrences of ice ages, as comprehensively reviewed in the past.\cite{Sr11, Sr12, Sr13, Sr14} The SR was first verified experimentally on an electronic circuit with a binary switch,\cite{Fauve83} but latter on many additional experimental findings were reported, as for example in a bistable non-linear optical device.\cite{McNam88}

Directly linked to the subject of the present work are studies focusing on noisy bistable oscillators, \cite{Jung91, Mor95a} as well as studies that consider coupled bistable systems \cite{Jung92, Buls93, Mor95b, Brey96, Siew98} within different settings of SR. The latter setup served also as a basis for the discovery of an interesting effect of noise on an ensemble of coupled bistable oscillators; namely the so-called array enhanced SR.\cite{Lind95, Lind96} Thereby, Lindner \textit{et al.} demonstrated that a linear coupling, combined with noise and a weak periodic signal, can enhance synchronization and global organization in a chain of overdamped nonlinear oscillators. Notably, investigations of bistable noisy systems remain vibrant to date, and in the last few years several theoretical \cite{Pasc03a, Pasc03b, Pasc05} as well as experimental \cite{Balt03, Khov07} studies have been performed that report on interesting new aspects of SR in such systems. Furthermore, noise-supported signal propagation \cite{Lindner98, GO00, Khov08} and the SR of collective variables with very large gains \cite{Casa06} in arrays of bistable systems have been studied as well. Recently, the scope of SR and related phenomena in coupled oscillators shifted also to ensembles characterized with complex interaction topologies, as constituted by small-world or scale-free networks.\cite{Gao01, Aceb07, Perc08}

Since soft matter systems \cite{SM} evolve both in time and space, and are in addition to that extremely susceptible to external perturbations, it is rather remarkable that they have avoided being the subject of investigations related to SR in the past. In particular so since during the last two decades the phenomenon of SR has been reported in the most diverse circumstances, ranging from semiconductor devices, \cite{kittel93} magnetoelastic ribbons, \cite{Spano92} superconducting resonators, \cite{Abdo07} intercellular calcium wave dynamics, \cite{Perc07} SQUID based detectors for weak magnetic fields, \cite{Rou95} to sociological models. \cite{Babi97} Moreover, the SR effect has also been reported in chemical reactions, where it has been shown that optimal noise intensities are able to enhance weak periodic \cite{Schneider1, Schneider2, Schneider3, Yang99} as well as aperiodic \cite{Showalter1} signals, or even support traveling waves in a spatially extended chemical medium.\cite{Showalter2} Recently, the popular nanoscale systems became associated with the SR as well. In particular, it has been shown that nanomechanical oscillators in a dynamic bistable state exhibit a more controllable switching in the presence of noise.\cite{Bad05}

At present, we aim to extend the scope of SR also to soft matter systems, considering a polymer stabilized ferroelectric (PSF) LC cell as a typical soft matter representative. Although the phenomenon of SR is most commonly associated with dynamic disorder or noise, soft matter systems, and indeed real-life systems in general, are in addition characterized by a certain degree of static disorder. This fact has recently been addressed by Tessone \textit{et al}., \cite{Tes06} who succeeded showing that static or quenched disorder can give rise to SR as well. Here we examine the impacts of both static and dynamic disorder on the phenomenon of SR, and demonstrate under which conditions it can be observed in a PSFLC cell. More precisely, we consider different origins of static disorder and show that a random field type static disorder yields qualitatively different responses with respect to random dilution, random bond and spin glass universality classes. In particular, while there always exists an optimal level of dynamic disorder, warranting the best correlation between a weak periodic electric field and the response of the PSFLC cell, the considered types of static disorder may additionally enhance or deteriorate the SR. Random dilution, random bond and spin glass universality classes decrease the level of dynamic disorder warranting the optimal response, whereas the random field evokes exactly the opposite effect, hence increasing the optimal level of dynamic disorder that is needed to resonantly fine-tune the system's response in accordance with the weak deterministic electric field. We additionally test these findings on their robustness with respect to the system size and range of interactions, and reveal that they are largely independent of such particularities. We discuss that, due to their extreme susceptibility to external perturbations and wide applicability, soft matter systems might represent an interesting environment for the development of SR-based sensitive detectors, and moreover, that additional theoretical as well as experimental research efforts in the direction outlined presently seem justified to realize these potentials.

The remainder of this paper is organized as follows. In Sec. II we present a generic dynamic equation giving rise to the SR phenomenon and the simulation procedure to solve it. Section III is devoted to the derivation of governing equations of the PSFLC cell, whereas in Sec. IV we present the results. In particular, we estimate analytically the influence of static disorder on the system's configuration, and numerically investigate the combined influence of static and dynamic disorder on the SR. The final section features the summary of presented results and outlines possible applications of our findings.

\section{Stochastic resonance: generic equations}

We consider a system of $N$ coupled bistable overdamped oscillators, governed by the Langevin equations of the form
\begin{equation}
\frac{\partial s_{i}}{\partial t}=
s_i-s_i^3+J \displaystyle \sum_{j}(s_j-s_i)+E+\sqrt{2D}\xi _{i}(t),
\label{gen}
\end{equation}
where $s_{i}$ describes the state of the $i$-th oscillator, $t$ is the dimensionless time, the sum index $j$ runs over the oscillators $s_{j}$ that are coupled with $s_{i}$, $J$ measures the coupling strength amongst the oscillators, $E = E_{0}\cos (\omega t)$ is a weak periodic field oscillating with the frequency $\omega$, whereas $2D$ is the variance of Gaussian noise with zero mean and autocorrelation  $\langle \xi_i (t) \xi_j (k) \rangle = \delta_{ij} \delta_{tk}$. Equation~(\ref{gen}) arguably provides a paradigmatic setup for different scenarios of SR, either via the classic setup, or via variations of system size \cite{Pik02} or diversity. \cite{Tes06} In the absence of deterministic forcing and noise each oscillator is characterized by a double-well potential with minima at $s_{i} = s_{0} = \pm 1$, whereas a periodic field $E$ with finite amplitude $E_{0} > 0$ modulates the double-well potential, potentially breaking the reflection symmetry of the system. In the regime of our interest $E_{0}$ is too weak to let the oscillators roll periodically from one potential well into the other. Instead, the minima are alternatively and asymmetrically pushed up and down, periodically raising and lowering the potential barrier. By increasing $D$ the noise-induced hopping between the potential wells can become synchronized with the frequency $\omega$ of the weak external field, thus exhibiting the SR phenomenon.

To evaluate responses of the noisy system to the periodic field $E$ in dependence on $D$, we calculate the Fourier coefficients for the mean field
\begin{equation}
S = \frac{1}{N}\sum_{i} s_{i}.
\label{mean}
\end{equation}
Noteworthy, the Fourier coefficients are proportional to the (square of the) spectral power amplification, \cite{fcx} which is often used as a measure for the quantification of SR. Here, the Fourier coefficients $Q$ are calculated according to the equations
\begin{eqnarray}
Q_{{\rm sin}} &=&\frac{\omega }{\pi n}\displaystyle \int_{0}^{2\pi n/\omega }S(t) \sin
(\omega t) {\rm d}t, \nonumber \\
Q_{{\rm cos}} &=&\frac{\omega }{\pi n}\displaystyle \int_{0}^{2\pi n/\omega }S(t) \cos
(\omega t) {\rm d}t, \label{Q} \\
Q &=& \sqrt{Q_{{\rm sin}}^2 + Q_{{\rm cos}}^2}, \nonumber
\end{eqnarray}
where $\omega $ is the frequency of the weak periodic field $E$ and $n$ is the number of oscillation periods used.

\section{Stochastic resonance in the PSFLC cell}

\begin{figure}
\centerline{\epsfig{file=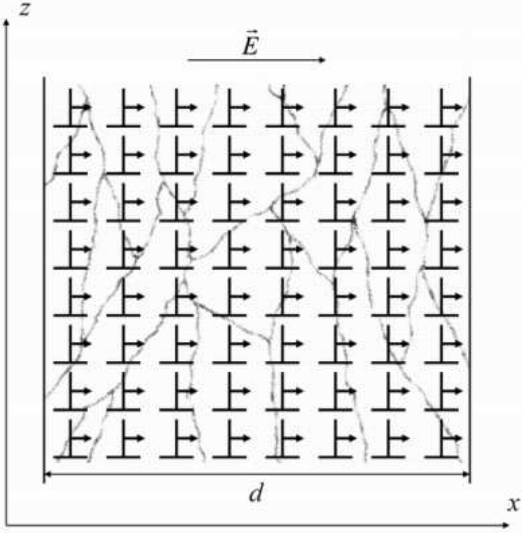, width=7cm}}
\caption{The SSFLC cell and the geometry of the problem. LC molecules are collected in smectic layers that are stacked along the $z$-axis. For $d < p$ the molecules are either tilted along the positive or the negative $y$-axis. Consequently $\vec{P}$ is aligned along the negative or the positive $x$-axis, respectively. The nail representation is used in order to show that LC molecules are tilted in the $(y, z)$ plane. Arrows attached to the `nails' mark the local polarization of molecules, whereas muddy gray lines in the background roughly indicate the structure of the polymer network. The weak periodic electric field $\vec{E}$ is introduced along the $x$-axis.}
\label{fig:fig1}
\end{figure}

In order to demonstrate an appropriate soft matter system that can potentially exhibit SR, we start by considering a thermotropic bistable surface stabilized ferroelectric (SSF) LC cell. \cite{Cla80} The ferroelectric LC, exhibiting both translational and orientational (quasi) long-range order, is confined in a plan-parallel cell as shown in Fig.~\ref{fig:fig1}. The translational order is characterized by smectic layers that are stacked along the $z$-axis, and the normals of the cell walls are parallel with the $x$-axis of the system. The orientation ordering is described by the director field $\vec{n}$, yielding an average local orientation of LC molecules, where $|\vec{n}| = 1$. The tilt (also referred to as the cone) angle $\theta = \arccos (\vec{n} \cdot \vec{v})$ between the smectic layer normals $\vec{n}$ and $\vec{v}$ distinguishes between the smectic A (SmA) and chiral smectic C (SmC*) ordering. The SmA ordering, in which $\langle \theta \rangle = 0$, is commonly realized above the critical temperature $T_{c}$, whereby $\langle \cdot \rangle$ stands for the spatial average. Below $T_{c}$ the LC enters into a helicoidal SmC* ordering. In a bulk sample $\langle \theta \rangle > 0$ the direction of the tilt precesses as one goes from one smectic layer to another. The resulting helicoidal structure is characterized by the wave number $q = 2 \pi /p$, where $p$ stands for the pitch of the helix. Tilting of the molecules gives rise to the in-plane spontaneous polarization $\vec{P}$, which is in general perpendicular to $\vec{n}$ and $\vec{v}$, whereby due to symmetry reasons it holds $\vec{P} = P \vec{v} \times \vec{n}$. \cite{Blinc91} Therefore, in the bulk SmC* phase $\vec{P}$ also forms a helix and the macroscopic polarization of the system equals zero. The cell confinement can suppress the helicoidal structure if the thickness $d$ of the cell is less or comparable to $p$. The resulting unwound SmC structure is on average homogeneously tilted along a single direction. In the case of isotropic planar anchoring (\textit{i.e.} molecules tend to lie in the confining plane within which all directions are equivalent), the molecules tilt either along the positive or the negative $y$-axis. We henceforth refer to these configurations as Up and Down states, which are separated by an energy barrier. Importantly, such a system thus exhibits bistability. The resulting polarization is aligned either along the positive or the negative $x$-axis, respectively. If a periodic external electric field is applied along the $x$-axis, it alternately favors the Up and Down configuration. If also a noisy component is present, then all the three essential conditions for the SR phenomenon listed in the Introduction are fulfilled.

We proceed by describing the bistable equilibrium structure in the unwound SmC phase quantitatively in terms of the tilt angle $\theta $ and the polarization $\vec{P}$. We consider the system depicted in Fig.~\ref{fig:fig1} where $p < d$. The molecules are either in the Up or Down state, \textit{i.e.} $\vec{n} = (0, |\sin \theta|, \cos \theta)$ and $\vec{P}=(\mp |P|, 0, 0)$. We further apply an external electric field $\vec{E} = E_{0}(\cos (\omega t), 0, 0)$. We expand the corresponding free energy density $f$ in terms of $\theta $ and $P$. Following Garoff and Meyer we obtain\cite{Gar77}
\begin{widetext}
\begin{equation}
\Delta f=f-f_{0}=\frac{\alpha _{0}(T-T_{0})\theta ^{2}}{2}+\frac{\beta
\theta ^{4}}{4}-CP\theta +\frac{2\theta P^{2}}{\chi \varepsilon _{0}}-EP+\frac{K}{2}\left\vert \nabla \theta \right\vert ^{2}+f_{w},
\label{f}
\end{equation}
\end{widetext}
where $f_{0}$ is the free energy density of the undistorted SmA phase, $\alpha _{0}$, $\beta$ and $T_{0}$ are the mean-field material constants, $C$ is the piezoelectric coupling constant, $\varepsilon _{0}$ is the dielectric constant, $\chi$ is the electric susceptibility, $K$ is the representative elastic constant of the SmC phase, and $f_{w}(\theta)$ describes the free energy costs at the LC-wall interface.

Minimization of $f$ with respect to $P$ yields $P = \varepsilon_{0} \chi (E+C \theta)$. We take this into account and introduce scaled dimensionless quantities $s= \theta \sqrt{\frac{\beta}{\alpha_{0}(T_{c}-T)}}$, $\widetilde{E}_{0}=E_{0}C \varepsilon _{0} \chi \left(\frac{\beta}{\alpha{0}(T_{c}-T)}\right) ^{3/4}$, $\widetilde{\Delta f}=\Delta f \frac{\beta}{\alpha_{0}^{2}(T_{c}-T)^{2}}$, $\widetilde{f_{w}} = f_{w} \frac{\beta}{\alpha_{0}^{2}(T_{c}-T)^{2}}$, $\widetilde{\nabla }=d \nabla$, and $J = \frac{K}{d^{2}} \frac{\beta }{\alpha _{0}^{2}(T_{c}-T)^{2}}$, where $T_{c}=\frac{C^{2} \varepsilon _{0} \chi }{\alpha _{0}}$ describes the second order structural phase transition temperature for $E = 0$. If omitting the tildes it follows $\Delta f = f_{c}+f_{e}+f_{f}+f_{w}$, where
\begin{eqnarray}
f_{c} &=&-\frac{s^{2}}{2}+\frac{s^{4}}{4},  \label{fc} \\
f_{e} &=&\frac{J}{2}\left\vert \nabla s\right\vert ^{2},  \label{fe} \\
f_{f} &=&-sE_{0}\cos (\omega t).
\label{ff}
\end{eqnarray}
The condensation term $f_{c}$ enforces $s=s_{0}= \pm 1$, the elastic term $f_{e}$ penalizes deviations from a spatially homogeneous ordering, and the term $f_{f}$ introduces the source of the coherent input signal. We proceed by adopting a standard form of the dissipation free energy term, neglect the $f_{w}$ contribution, and after discretization we reproduce Eq.~(\ref{gen}), where $s_{i}=s(\vec{r}_{i})$. Finally, in a SSFLC cell, the dynamic Gaussian noise $\xi_{i}(t)$ entering Eq.~(\ref{gen}) could be produced by a randomly varying external electric field.

Next, we rewrite Eq.~(\ref{gen}) into a more general form
\begin{equation}
\frac{\partial s_{i}}{\partial t}=s_{i}-s_{i}^{3}+ \displaystyle \sum_{j}J_{ij}(s_{j}-s_{i})+E+\sqrt{2D}\xi_{i}(t)+s_{i}w_{i},
\label{dyn1}
\end{equation}
where the quantities $J_{ij}$ and $w_{i}$ now introduce qualitatively different origins of static disorder. The physical origin of these contributions in systems of our interest is as follows. Standard SSFLC cells are extremely susceptible to imperfections of the confining walls. One can substantially improve their mechanical stability by introducing a polymer network into the LC. The resulting system is referred to as the PSFLC. For a low enough concentration of the polymer the bistability is preserved. The network introduces a kind of a bulk field. Its local impact could be approximately modeled by the LC-polymer coupling term $f_{w} \sim - \frac{w_{i}s_{i}^{2}}{2}$, giving rise to the $s_{i}w_{i}$ contribution in Eq.~(\ref{dyn1}). This term introduces, due to the essentially random nature of the polymer network, a certain degree of static randomness or disorder into the system.\cite{Pop1, Pop2} Furthermore, the presence of the polymer can give rise to spatially varying elastic interactions within the system, resulting in spatially dependent coupling constants $J_{ij}$. Lastly, $E = E_{0}\cos (\omega t)$ is now a generalized weak periodic field with a dynamic random component approximated by the Gaussian noise $\xi_{i}(t)$ with the same properties as in Eq.~(\ref{gen}).

Prior to examining the results, we further generalize Eq.~(\ref{dyn1}) in order to investigate a variety of soft matter systems experiencing qualitatively different origins of disorder. We consider influences of random dilution (RD), random bond (RB), spin glass (SG), and random field (RF) type of static disorder. By the RD type we set $J_{ij} = 0$ at randomly chosen interactions with a probability $\gamma$, whereas at the remaining sites it holds $J_{ij} = J$. By the RB type we allow $J_{ij}$ to vary randomly within the interval $J_{ij} \in [J - \Delta J, J]$, where $0 < \Delta J < 2J$. The case $\Delta J = 2J$ corresponds to the SG universality class, where $J_{ij}$ exhibits random values within the interval $[-J, J]$. By the RF type of static disorder a value of the field $w_{i}$ is chosen randomly within the interval $[-\Delta w,\Delta w]$. Furthermore, we consider either short-range (interactions only amongst the four nearest neighbors) or infinite-range (all-to-all coupling) interactions $J_{ij}$. In the latter case we assume that a value of $J_{ij}$ is independent of the distance between the $i$-th and $j$-th oscillator.

\section{Results}

In what follows, we study the phenomenon of SR in the system described by Eq.~(\ref{dyn1}). We first focus on the influence of different types of static disorder via an analytical treatment of equilibrium values of the $i$-th oscillator $s_{i}$ within the ensemble. Afterwards, we study numerically the combined influence of dynamic and static disorder on the mean-field response of the studied soft matter system.

\subsection{Influence of static disorder}

We estimate how different sources of static disorder modify the bistable configuration enforced by the condensation term $f_{c}$. Since bistability is a key ingredient enabling the observation of SR, these insights will help us to understand and interpret effects of the joint impact of static and dynamic disorder that we are going to present in the next subsection. In order to assert the impact of static disorder on bistability, we treat disorder terms (\textit{i.e.} the $J_{ij}$ and $w_{i}$ contribution) as weak perturbations. Moreover, we neglect the dynamic disorder [Gaussian noise $\xi_{i}(t)$], switch off the periodic electric field $E$, and express $s_{i}$ as
\begin{equation}
s_{i}=s_{0}+x_{i},
\label{xi}
\end{equation}
where $s_{0} = \pm 1$. Linearization of the statical part of Eq.~(\ref{dyn1}) with respect to $x_{i}$ yields $-2x_{i}+\sum_{j} J_{ij}(x_{j}-x_{i})+w_{i} (s_{0}+x_{i}) = 0$. For $J_{ij}=0$ we obtain in the lowest approximation
\begin{equation}
x_{i} \sim s_{0}w_{i}/2.
\label{d1}
\end{equation}
Accordingly, we conclude that the presence of RF type disorder of any strength modifies the equilibrium configuration of $s_{i}$.

To estimate the influence of random variations in $J_{ij}$ we set $w_{i}=0$. For convenience we introduce the renormalized random interaction coupling matrix $J_{ij}^{\prime}$ as
\begin{equation}
J_{ij}^{\prime }=J_{ij}-\delta _{ij}\sum_{k}J_{kj},
\label{Jij}
\end{equation}
where $\delta _{ij}$ is the Kronecker symbol. We further express $x_{i}$ in the eigenbasis \cite{Her83,Cho84} of the matrix $J_{ij}^{\prime}$
\begin{eqnarray}
x_{i} =\sum_{\alpha }A_{\alpha }e_{i}^{(\alpha )},  \label{e1a} \\
\sum_{j}J_{ij}^{\prime }e_{j}^{(\alpha )} = \Omega_{\alpha }e_{i}^{(\alpha)},
\label{e1b}
\end{eqnarray}
where $\Omega_{\alpha }$ are the eigenvalues, $e_{i}^{(\alpha )}$ are the corresponding eigenvectors and $A_{\alpha }$ are the weight constants. The eigenvectors are normalized according to
\begin{equation}
\sum_{i}e_{i}^{(\alpha )}e_{i}^{(\beta )}=\delta _{\alpha \beta}.
\label{en}
\end{equation}
We expand the statical part of Eq.~(\ref{dyn1}) for $E=0$ including the cubic term in $x_{i}$, take into account Eqs.~(\ref{e1a}) and (\ref{e1b}), multiply the resulting equation with $e_{i}^{(\eta)}$, and sum it over the lattice sites. It follows
\begin{widetext}
\begin{equation}
\sum_{\alpha ,\beta ,\gamma }A_{\alpha }A_{\beta }A_{\gamma }F_{4}^{(\alpha
\beta \gamma \eta )}+3\sum_{\alpha ,\beta }A_{\alpha }A_{\beta}F_{3}^{(\alpha \beta \eta )}+A_{\eta }(\Omega -\Omega_{\eta })=0,
\label{Eqm1}
\end{equation}
\end{widetext}
where
\begin{equation}
F_{4}^{(\alpha \beta \gamma \eta )}=\sum_{i}e_{i}^{(\alpha)}e_{i}^{(\beta)}e_{i}^{(\gamma)}e_{i}^{(\eta )}, F_{3}^{(\alpha \beta \eta)}=\sum_{i}e_{i}^{(\alpha )}e_{i}^{(\beta )}e_{i}^{(\eta )}.
\label{Fi}
\end{equation}
Neglecting the coupling \cite{Cho84} among different eigenmodes  and assuming $F_{3}^{(\eta )}=0$ one obtains the equation for the $\eta$-th mode amplitude
\begin{equation}
A_{\eta }^{3}F_{4}^{(\eta )}+A_{\eta }(\Omega -\Omega_{\eta })=0,
\label{Eqm2}
\end{equation}
where $\Omega=2$.
It follows $A_{\eta }(\Omega_{\eta }< \Omega )=0$ and
\begin{equation}
A_{\eta }(\Omega_{\eta }>\Omega )=\sqrt{\frac{\Omega_{\eta }-\Omega }{F_{4}^{(\eta )}}}.  \label{Am}
\end{equation}
The corresponding effective Hamiltonian $H_{\eta }$ determining the $\eta $-th eigenmode amplitude can be defined as
\begin{equation}
H_{\eta }=\frac{A_{\eta }^{2}}{2}(\Omega -\Omega_{\eta}) + \frac{A_{\eta}^{4}}{4}F_{4}^{(\eta)},
\label{Hm}
\end{equation}
where $\Omega =2$ plays the role of the generalized temperature. On decreasing $\Omega$ the eigenmodes open via a continuous phase transition at $\Omega =\Omega_{\eta }$ and below $\Omega_{\eta }$ its amplitude is given by Eq.~(\ref{Am}). It follows that the disorder in $J_{ij}$ can trigger structural changes only for large enough disorder strengths, \textit{i.e.} at least one eigenvalue must be larger than $\Omega $. When departing from the bistable solution preferred by $f_{c}$ only modes characterized by $\Omega_{\eta } > \Omega$ are thus present.

Note that in the derivation of Eq.~(\ref{Am}) we have assumed that the eigenmodes do not overlap. This is partly justified only for localized modes \cite{Cho84}. These are expected for short-range interactions in $J_{ij}$ and $\Omega_{\eta }>0$. On average they appear at different regions and consequently avoid overlapping. The response of the system is in this case given by
\begin{equation}
s_{i}=s_{0}+\sum_{\alpha} A_{\alpha } e_{i}^{(\alpha)},
\end{equation}
where only the modes with $\Omega_{\alpha } > \Omega =2$ participate. The extended modes thus appear only for $\Omega_{\eta } \sim 0$. \cite{Metha67}

In case of long-range interactions in $J_{ij}$ the modes are extended. \cite{And78} For example, for infinite-range interactions and SG type disorder the distribution of eigenvalues stays within the interval $\Omega_{\eta} \in [-\Omega_{\max}, \Omega_{\max }]$ according to the symmetric Wigner probability distribution \cite{Metha67}
\begin{equation}
P(\Omega_{\eta })=\frac{2}{\pi \Omega_{\max}}\sqrt{\Omega_{\max}^{2}-\Omega_{\eta}^{2}},
\label{PSG}
\end{equation}
where $\Omega _{\max }$ stands for the maximal eigenvalue. Therefore, a single mode is expected to dominate in $x_{i}$, and this mode hinders the opening of the remaining modes. Further, if $F_{3}^{(\eta )} \neq 0$ then the modes would appear via a discontinuous transitions as a function of $\Omega$.

To illustrate the influence of static disorder and different interaction ranges on $x_{i}$ configurations we focus on the quantity $1/F_{4}^{(\eta )}$, where $F_{4}^{(\eta )}$ is defined by Eq. (\ref{Fi}). It estimates the number $N_{\eta }$ of sites participating in the $\eta$-th mode, \textit{i.e.} $1/F_{4}^{(\eta )}\sim N_{\eta }$. The estimate follows from the normalization condition given by Eq. (\ref{en}). We assume that all `open' sites, characterized by $x_{i}^{(\eta)}\neq 0$, are comparable in magnitude. Therefore $|x_{i}^{(\eta )}| \sim 1/a$ if site $i$ is opened, whereby $a$ stands for a positive constant, but otherwise $x_{i}^{(\eta )}=0$. In Eq. (\ref{en}) the sum is restricted over $N_{\eta }$ open sites, resulting in $N_{\eta}/a^{2}=1$. Consequently, $|x_{i}^{(\eta)}| \sim 1/\sqrt{N_{\eta }}$ and $F_{4}^{(\eta )}= \sum_{i} \left(e_{i}^{(\eta )}\right) ^{4}\sim $ $1/N_{\eta}$. Non-localized (extended) modes are characterized by $N_{\eta } \sim N$ and localized by $N_{\eta }\ll N$.

\begin{figure*}
\centering
\includegraphics[width=13cm]{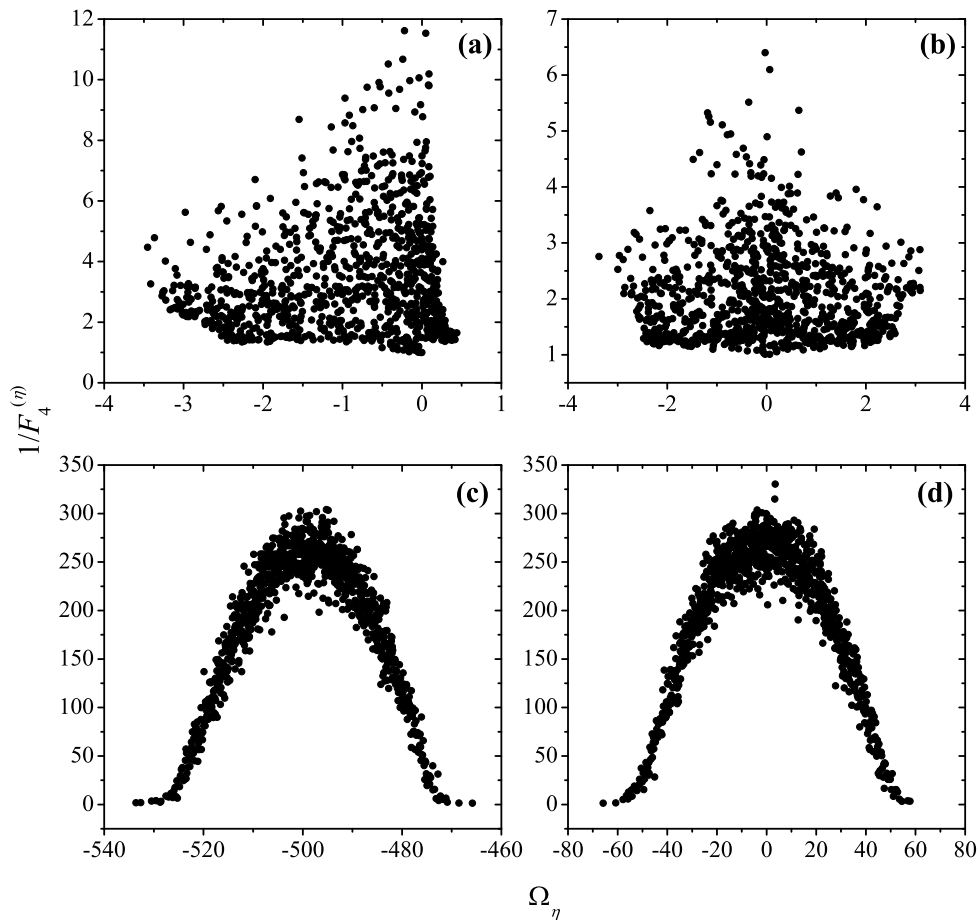}
\caption{The approximate mode size $\frac{1}{F_{4}^{(\eta )}}$ as a function of $\Omega_{\eta}$. Type of disorder and range of interactions are as follows: (a) RB, $\Delta J=J$, short-range; (b) SG, $\Delta J=2J$, short-range; (c) RB, $\Delta J=J$, infinite-range; (d) SG, $\Delta J=2J$, infinite-range. System size in all panels was $N=1000$.}
\label{fig:fig2}
\end{figure*}

In Fig.~\ref{fig:fig2} we plot $1/F_{4}^{(\eta )}$ $\sim N_{\eta }$ as a function of mode eigenvalues $\Omega _{\eta }$ for $N=1000$. We consider the RB case with $\Delta J=J$, and the SG limit $\Delta J=2J$. The resulting $1/F_{4}^{(\eta )}$ values are shown for short- and infinite-range interactions in the top and bottom panels, respectively. By short-range interactions the modes are typically localized [in Figs.~\ref{fig:fig2}(a) and (b) $N_{\eta }$ is typically less than 10]. On the contrary, by infinite-range interactions most states are extended [characterized by $N_{\eta}>100$, as shown in Figs.~\ref{fig:fig2}(c) and (d)]. We further find that by increasing the value of $\Delta J$, where $J_{ij}\in [J-\Delta J,J]$, the maximal value $\Omega_{\max }$ increases. According to the estimate given by Eq. (\ref{Fi}) the modes can open if $\Omega_{\max} > \Omega =2$. Our calculations suggest that such modes are enabled only for $\Delta J>J$. Consequently, we find that the disorder in $J_{ij}$ notably affects the SR phenomenon only if $\Delta J>J$. Note that Eq. (\ref{Fi}) holds only for relatively weak static disorders via $J_{ij}$, and for the case that modes do not strongly overlap.

We further show that the disorder introduced via the coupling interaction $J_{ij}$ is unlikely to destroy the bistability enforced by the condensation term. For this purpose we originate from Eq.~(\ref{dyn1}) and set $E=0$, $w_{i}=0$, and $\sigma _{i}=0$. We express $s_{i}$ as
\begin{equation}
s_{i}=S+\delta s_{i},
\label{Si}
\end{equation}
where $\delta s_{i}$ describes fluctuations around the mean field $S$ of the system. If the bistability is preserved, then $S=\pm | S |$. In the opposite case $S=0$. We sum the resulting equation over all lattice sites, assume $\frac{1}{N}\sum_{i} \frac{\partial \delta s_{i}}{\partial t}\sim \frac{1}{N}\sum_{i} \delta s_{i}\sim \frac{1}{N} \sum_{i} \delta s_{i}^{3}\sim 0$, and obtain the equation
\begin{equation}
S-S^{3} - 3 S M=0,
\label{Sm}
\end{equation}
where $M$ measures the strength of departures about the average:
\begin{equation}
M=\frac{1}{N}\sum_{i} \delta s_{i}^{2}.
\label{M}
\end{equation}
It follows that the bistability is preserved for $M<1/3$, where
\begin{equation}
S=\pm \sqrt{1-3M}.
\label{Sm1}
\end{equation}

Finally, within this subsection we address the condition where the bistability is lost with eigenmodes of the matrix $J_{ij}^{\prime}$. From Eqs.~(\ref{Si}) and (\ref{xi}) it follows $\delta s_{i}=s_{0}-S+x_{i}$. The bistability is lost when $M=1/3$ and $S=0$. With this in mind, using the definition given by Eq.~(\ref{M}) and expansion given by Eq.~(\ref{e1a}), we obtain the condition
\begin{equation}
\frac{1}{N}\sum_{m} A_{m}^{2}+\frac{2S_{0}}{N}\sum_{m} A_{m}F_{1}^{(m)}\sim \frac{1}{N}\sum_{m} A_{m}^{2}=-2/3,
\label{bi}
\end{equation}
where the sum runs over the open modes, the amplitudes of which are estimated by Eq.~(\ref{Am}). It is obvious that the condition given by Eq.~(\ref{bi}) cannot be satisfied. Therefore, the static disorder in the coupling strength cannot, on its own, destroy the bistability.

\subsection{Combined influence of static and dynamic disorder}

\begin{figure}
\centering
\includegraphics[width=7cm]{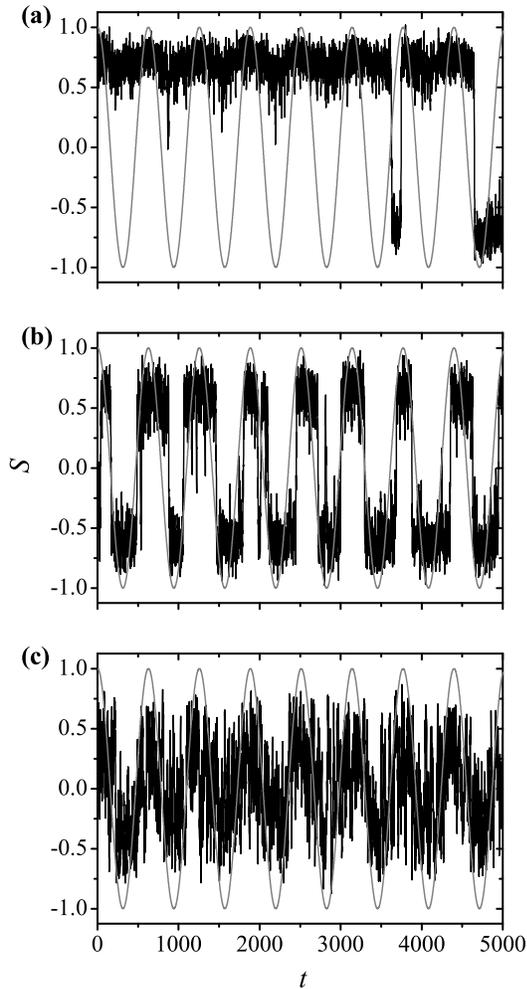}
\caption{Temporal evolution of the mean-field response $S$ of the system by different levels of (dynamic) Gaussian noise $D$. (a) $D=0.7$, (b) $D=0.9$ and (c) $D=1.3$. Gray lines depict the periodic driving (amplitude is scaled to 1). In panel (b) the resonance condition is fulfilled, resulting in the optimal correlation between the weak periodic driving and the response of the system. Employed system parameters are: $J_{ij}=J=4$, $w_{i}=0$, $E_{0}=0.04$, $\omega=0.01$ and $N=100$.}
\label{fig:fig3}
\end{figure}

\begin{figure}
\centering
\includegraphics[width=7cm]{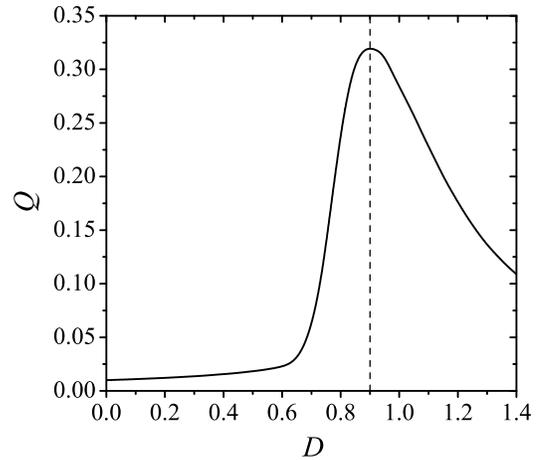}
\caption{SR in the examined system. By an intermediate value of $D$ (denoted by the dashed vertical line) the correlation of the mean-field response $S$ with the weak periodic driving $E$ is maximal, resulting in a bell-shaped dependence of $Q$. System parameters were the same as in Fig.~\ref{fig:fig3}.}
\label{fig:figx}
\end{figure}

In this subsection we analyze numerically the response of the system in terms of the mean field $S$ if both static and dynamic disorder are introduced. We use periodic boundary conditions for our simulations and consider short- and infinite-range interactions in $J_{ij}$. In the former case only the nearest neighbors are coupled, whereas in the latter the coupling is independent of the physical distance between the units. By considering both short- and infinite-range couplings we take into account two limiting cases of interactions that interpolate between the 3D structures usually entailed within soft matter systems. Importantly though, we found the actual usage of 3D models prohibitive due to substantial computer resources that would be needed to simulate them. We do, however, demonstrate that our results are independent of the system size in Fig.~\ref{fig:fig5}, and thus should be readily observed also in the thermodynamic limit.

We start by examining the temporal evolution of the mean-field response $S$ for three different levels of Gaussian noise $D$. As the reference state we consider a configuration with $J_{ij}=J=4$ in the absence of any additional static disorder and infinite-range of interactions. Figure~\ref{fig:fig3} features the results. In panel (a) the noise level is clearly to weak to assist the periodic field strong enough to induce flips between the two minima of the potential (note that $E_{0}=0.04$ assures that $E$ itself is too weak to induce the flips). In sharp contrast, panel (c) depicts a heavily noisy response, lacking a clear temporal structure that would indicate any particular correlations with the weak deterministic driving. Only in panel (b) the correlation between $S$ and the weak periodic field $E$ is evident already with the naked eye, and indeed, visually manifests the SR phenomenon in the examined soft matter system. To support this visual assessment, we examine $S$ evoked by different $D$ via the Fourier coefficients $Q$ [see Eq.~(\ref{Q})] that quantify the correlation between $S$ and $E$. Figure~\ref{fig:figx} depicts a typical bell-shaped dependence of $Q$, thus confirming that an intermediate level of Gaussian noise is able to optimally assist the weak periodic driving $E$ to induce flips between the two minima of the potential. This is the hallmark of the SR effect, which we thus demonstrate in the examined soft matter system.

\begin{figure*}
\centering
\includegraphics[width=15cm]{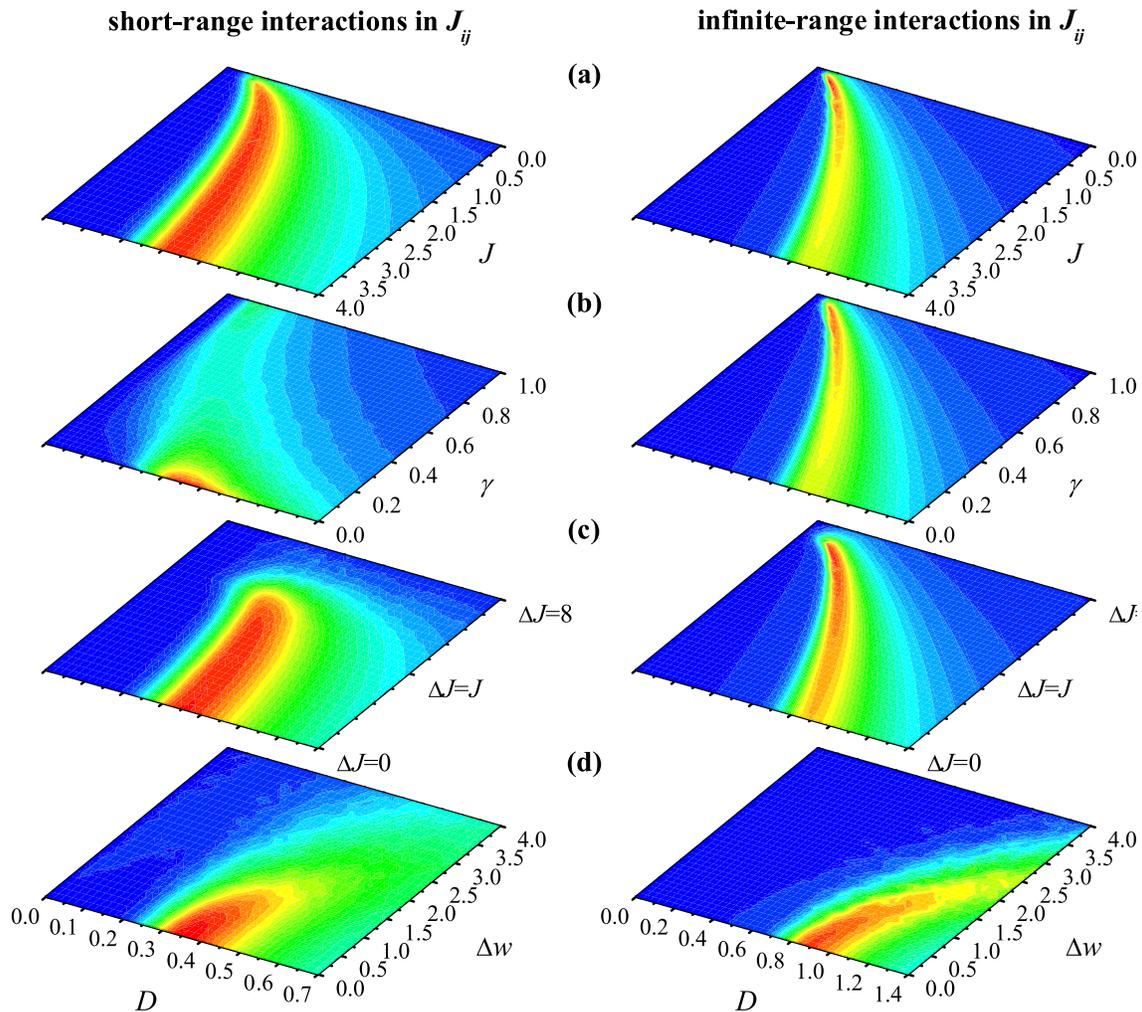}
\caption{Response of the system, quantified via $Q$ [see Eq.~(\ref{Q})], in dependence on the Gaussian noise level $D$ as well as the type and strength of static disorder. Results are presented separately for short-range (left column) and infinite-range interactions (right column). Rows depict: (a) impact of different $J_{ij}=J$; (b) RD; (c) RB; and (d) RF. In all panels the color profile is RGB linear (blue $>$ green $>$ yellow $>$ red), blue depicting $0.0$ and red $0.45$ values of $Q$. The employed system size was $N=100$ and $Q$ were evaluated over $n = 1000$ oscillation periods of $E$. In addition, all results were averaged over 50 different realization of every type of static disorder. Where applicable, other parameters were the same as in Fig.~\ref{fig:fig3}.}
\label{fig:fig4}
\end{figure*}

In order to distinguish impacts of different types of static disorder on the SR resonance effect exemplified in Fig.~\ref{fig:figx}, we continue with examining the mean field responses $S$ via the Fourier coefficients $Q$. First, we consider the influence of decreasing coupling strengths $J_{ij}=J$ to obtain a reference for further simulations. In particular, note that both the RD and RB types of static disorder progressively lower the average coupling strength $\overline{J_{ij}}$ as their magnitude increases, and thus the reference simulation in which the interaction strength between all the coupled oscillators $J$ is gradually reduced may provide important pointers for the understanding of the effect of RD and RB static disorder. The resulting responses quantified via $Q$ in dependence on $D$ and $J$ are shown in Fig.~\ref{fig:fig4}(a) separately for short- and infinite-range interactions. For $J=4$ the correlation of $S$ with $E$ peaks at $D=D_{r} \approx 0.35$ (short-range) and $D_{r} \approx 0.9$ (infinite-range; see also Fig.~\ref{fig:figx}), whereby the subscript $r$ marks the resonance value of $D$. In both cases the resonance peak shifts toward lower values of $D=D_{r}$ as the coupling strength is reduced, and moreover, an accompanying increase in peak values of $Q$ prior to reaching $J=0$ can be inferred in case of infinite-range interactions (the global maximum of $Q$ is obtained at $D=D_{r}=0.11$ and $J=0.17$). At a glance, it can be observed that RD [Fig.~\ref{fig:fig4}(b)] and RB [Fig.~\ref{fig:fig4}(c)] types of static disorder by infinite-range interactions (right column) deliver qualitatively identical results as obtained by the reference simulation [Fig.~\ref{fig:fig4}(a)]. Indeed, by increasing $\gamma$ (RD) or $\Delta J$ (RB) the average coupling strength $\overline{J_{ij}}$ decreases, which due to the fact that by infinite-range interactions each unit experiences approximately the same coupling with others, simply results in the same effect as if $J$ would be gradually reduced equally for all units. Accordingly, in the right (b) and (c) panels the optimal $D=D_{r}$ decreases as $\gamma$ and $\Delta J$ increase, and global maxima (across the whole 2D parameter space) of $Q$ are obtained prior to reaching the limits of the RD and RB universality class. Specifically, the global maxima of $Q$ are obtained at $D=D_{r}=0.11$ and $\gamma=0.96$ by the RD case, and by $D=D_{r}=0.18$ and $\Delta J=7.1$ by the RB case. Conversely, by short-range interactions the individuality of couplings amongst units is more pronounced, and accordingly, the reference simulation in the left column of Fig.~\ref{fig:fig4}(a) bears less similarity with the impact triggered by RD and RB types of static disorder, especially if compared to infinite-range interactions. In case of short-range interactions the RD and RB types of static disorder introduce effects that go beyond a simple decrease in $J$. The main trend, however, remains the same in that the resonance peak of $Q$ shifts towards lower values of $D = D_{r}$ as the magnitude of static disorder increases, but nevertheless, the global maxima of $Q$ are obtained by $\gamma = 0$ and $\Delta J = 0$. Moreover, substantial removal of links with probability $\gamma$ by the RD case seems to have a rather profoundly deteriorating effect on the maximally attainable peak value of $Q$.

\begin{figure}
\centering
\includegraphics[width=7cm]{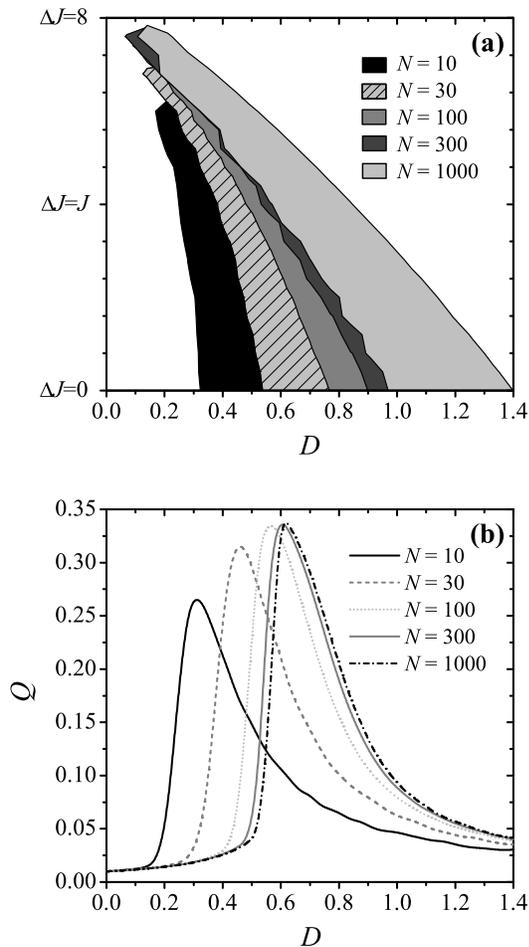}
\caption{Joint effects of RB type static disorder and Gaussian noise in dependence on the system size $N$ by infinite-range interactions. (a) Shaded outlines, representing $Q > 0.18$, of the resonance behavior for different $N$. (b) Excerpts of SR depicted in panel (a) obtained by $\Delta J = J$. Importantly, the SR is preserved and saturates in magnitude above $N \sim 300$.}
\label{fig:fig5}
\end{figure}

Most notable, however, is the qualitatively different response that is evoked by the RF type of static disorder depicted in Fig.~\ref{fig:fig4}(d), which sets in irrespective of the range of interactions since $w_{i}$ always has only a localized impact on each particular unit. It can be inferred that by the RF type of static disorder the resonance peak shifts toward higher values of $D = D_{r}$ as $\Delta w$ increases, and also, an accompanying decrease in peak values of $Q$ can be inferred (global maxima of $Q$ are obtained at $\Delta w = 0$). This is significantly different from what we have observed by RD and RB types of static disorder, especially so by infinite-range interactions. The discrepancy can be explained by our preceding separate treatment of effects of static disorder. In particular, unlike variations in $w_{i}$, disorder introduced via $J_{ij}$ does not change the double well potential of $s_{i}$ if $s_{i}=s_{j}$ since such pairs are preferred for a positive value of $J_{ij}$, and hence the essential bistability property for SR remains intact. On the other hand, disorder introduced via $w_{i}$ directly influences the energy landscape experienced by $s_{i}$, and thus destroys bistability for large enough $\Delta w$. This in turn explains the ever increasing values of $D = D_{r}$ needed for a resonant response and the deteriorating peak values of $Q$ depicted in both panels of Fig.~\ref{fig:fig4}(d) as $\Delta w$ increases.

Finally, we examine whether above results are qualitatively independent of the system size. The latter is an important issue since by soft matter systems the thermodynamic limit $N \rightarrow \infty$ can in general be considered fulfilled. For the purpose of this analysis we consider the RB type of static disorder as a representative example (for other types of static disorder identical results are obtained), and moreover, focus only on infinite-range interactions as variations in system size do not influence the SR in case of short-range interactions. Results for the full range of $\Delta J$ are presented in Fig.~\ref{fig:fig5}(a). Shaded are areas in the $D - \Delta J$ parameter space where $Q > 0.18$. Clearly, the SR is preserved as $N$ increases. To enable a better quantitative insight of the preservation of SR, we show in Fig.~\ref{fig:fig5}(b) excerpts of the resonant behavior for $\Delta J = J$. Indeed, it can be observed nicely that not only the combined effects of static and dynamic disorder are preserved, but also that the resonance behavior saturates for large enough $N$. This is in accordance with previously reported results of Monte Carlo simulations on a simple Ising-type lattice model, \cite{Ned95} thus fully confirming the generality of our findings with respect to variations in system size.

\section{Summary}

We studied the phenomenon of SR via variations of static and dynamic disorder in a soft matter system. As a representative example, we choose a polymer stabilized ferroelectric LC cell, which is an essential ingredient in several electro-optic applications.  The cell's thickness was considered shorter than the pitch of the helix, which resulted in a bistable dynamics governing the discretized equations of the LC configuration in the SmC phase with qualitatively identical properties as exhibited by the paradigmatic bistable overdamped oscillator. As the latter is a generic model for SR, the observation of noise-enhanced correlation between a weak periodic field and the PSFLC response could thus be expected. Indeed, by applying a subthreshold electric field $E$ along the surface normal of the two plates that alternatively favored the competing Up and Down configurations, we were able to demonstrate the phenomenon of SR in a typical soft matter system by measuring its mean-field response $S$.

More precisely, we focused on the combined influence of static and dynamic disorder, whereby the former arises naturally due to the random character of the polymer network that is introduced into the LC in order to improve its mechanical stability. In particular, the coupling between LC molecules can then no longer be considered uniform because of induced local variations in the effective temperature within the system, thus introducing RD and RB types of static disorder, and because of the branched polymer structure introducing RF type disorder. Importantly, we generalized the properties of the examined LC in order to obtain broader insights into the behavior of qualitatively different systems, which are however, governed by the same generic equations. In addition to considering these three types of static disorder, we also examined the system's dynamics in dependence on the system size as well as on local- and infinite-range interactions. First, we showed analytically that static disorders introduced via the coupling strength $J_{ij}$ (RD and RB) and via the random field $w_{i}$ (RF) affect the system's bistability in qualitatively different ways. While the RF static disorder is very effective in destroying the bistable character of the system even by small magnitudes, the RD and RB types of static disorder fail to have effects unless the level of randomness is large. We further show that the range of interactions strongly affects the response of the system with RD or RB type of disorder. These analytical estimates were obtained by taking into account a system with no dynamic disorder and relatively weak static disorder. Our numerical simulations, where the combined influence of static and dynamic disorder was taken into account, confirmed our analytical estimates. While RD and RB types of static disorder can be simplified mostly to decreasing coupling strengths $J_{ij}=J$ (with some differences emerging with respect to the range of interactions amongst LC molecules), the RF type disorder induces a qualitatively different response. Particularly, RD and RB (becomes SG in the maximal disorder limit) decrease the level of dynamic disorder warranting the optimal response, whereas the RF evokes exactly the opposite effect by increasing the optimal $D$ that is needed to resonantly fine-tune $S$ in accordance with $E$. These observations are shown to be independent of the system size and largely independent also of the range of interactions, which implies that they are generally valid and potentially applicable beyond the presently employed setup. Indeed, our study leads us to conclude that soft matter systems are viable candidates for additional theoretical as well as experimental research, potentially leading to sensitive detectors with fortified capabilities due to the phenomenon of SR.

\begin{acknowledgments}
Matja{\v z} Perc and Samo Kralj acknowledge support from the Slovenian Research Agency (Grants Z1-9629 and J1-0155).
\end{acknowledgments}

\end{document}